\documentclass[a4,12pt]{article}
\usepackage[dvips]{graphicx}
\usepackage{amsmath,amssymb}
\usepackage{amssymb}
\usepackage{enumerate}
\usepackage[usenames]{color}

\renewcommand{\thefootnote}{\fnsymbol{footnote}}

\begin{document}

\title{Non-Abelian Global Strings at Chiral Phase Transition}

\maketitle

\begin{center}
\author{
Muneto Nitta$^a$\footnote{e-mail: {\tt nitta(at)phys-h.keio.ac.jp}}
 and 
Noriko Shiiki$^{b,c}$\footnote{e-mail: {\tt shiiki(at)laputa.c.u-tokyo.ac.jp}},
}

$^a$ {\it Department of Physics, Keio University, Hiyoshi, Yokohama,
Kanagawa 223-8521, Japan}\\
$^b$ {\it Graduate School of Art and Science, 
University of Tokyo,
Komaba 3-8-1, Tokyo 153-8902, Japan}\\
$^c$ {\it Department of Management, Atomi University, 
Niiza, Saitama 352-8501, Japan}\\

\abstract
{
We construct non-Abelian global string solutions 
in the $U(N)_{\rm L} \times U(N)_{\rm R}$ linear sigma model.
These strings are the most fundamental objects 
which are expected to form during the chiral phase transitions, 
because the Abelian $\eta'$ string is marginally decomposed 
into $N$ of them. 
We point out Nambu-Goldstone modes of ${\bf C}P^{N-1}$ for 
breaking of $U(N)_{\rm V}$ arise around a non-Abelian vortex. 
}

\end{center}

\setcounter{footnote}{0}
\renewcommand{\thefootnote}{\arabic{footnote}}

\newpage


\section{Introduction}

Formation of topological defects 
is inevitable when a symmetry is spontaneously broken 
at a phase transition. 
Especially strings \cite{VS} are often formed 
in various area of physics.
They appear in 
the standard model \cite{Achucarro:1999it}, 
(SUSY) GUTs \cite{Jeannerot:2003qv} and other models 
in particle physics, and  
superconductors, superfluids, nematic liquid crystals, 
cold atom and so on in condensed matter physics. 
In cosmology they 
are proposed as cosmic strings \cite{VS}. 
Depending on whether 
a broken symmetry is global, local (gauged) 
or both, 
strings are called local, global, or semi-local. 
Local and global $U(1)$ (namely Abelian) strings 
were extensively studied in cosmology 
because these cosmic strings were proposed to 
play a role in structure formation \cite{VS}. 
Although this possibility was rejected once,  
recently there have been 
a revival of cosmic strings from superstring theory, 
brane inflation and possible observation 
of these strings \cite{recent}.
More recently non-Abelian local strings 
have been found in superstring theory \cite{Hanany:2003hp} 
and in supersymmetric QCD \cite{Auzzi:2003fs}.
These strings are BPS {\it i.e.} at the critical coupling, 
and so the most generic solutions and their moduli space 
have been obtained \cite{Eto:2005yh} 
in the moduli matrix formalism \cite{Isozumi:2004vg}.  
These strings have been proposed as non-Abelian cosmic strings; 
the reconnection (intercommutation) rate has been shown to be unity 
\cite{Hashimoto:2005hi}, 
and gravity coupling \cite{Aldrovandi:2007bn}
and thermal effects \cite{Eto:2007aw}
have been investigated.
Furthermore
non-Abelian {\it semi-local} strings 
have also been extensively studied 
\cite{Isozumi:2004vg,Shifman:2006kd}.

Another kind of non-Abelian strings not 
yet studied much is 
a non-Abelian {\it global} string.  
Non-Abelian global strings are not only theoretically 
interesting as a candidate for cosmic strings, but they 
might have been formed during the chiral phase transition of 
QCD where the chiral symmetry 
$U(N)_{\rm L} \times U(N)_{\rm R}$
is spontaneously broken to 
its diagonal subgroup $U(N)_{\rm V}$.
Strings at this  
transition were first discussed by 
Brandenberger et al \cite{Zhang:1997is}--\cite{Mao:2004ym}
where two types of global strings were proposed.
One, the pion string, is made of the pions
associated to broken $SU(N)_{\rm A}$ symmetry
and is topologically unstable\footnote{
Similar unstable string exists in the Skyrme model 
\cite{Jackson:1988xk}.
}.  
The other, an {\it Abelian} string called 
the $\eta'$ string, is made of the $\eta'$ meson
associated to broken $U(1)_{\rm A}$ symmetry
and is topologically stable 
at least at high temperature 
where the anomaly term disappears \cite{Pisarski:1983ms}.
At low temperature the anomaly term arises, 
the $\eta'$ meson acquires mass \cite{'t Hooft:1976up}, 
and the string is attached by domain walls 
\cite{Balachandran:2001qn} as in the case of axion strings \cite{VS}. 
The existence of {\it non-Abelian} global strings 
was first pointed out by Balachandran and Digal 
in a seminal paper \cite{Balachandran:2002je}\footnote{  
Non-Abelian {\it global} strings with $SU(N)_{\rm L}$ gauged 
appear in high density QCD \cite{Balachandran:2005ev}.
Similar global strings 
appear in the B-phase of ${}^3$He superfluids, 
in which chiral symmetry is replaced by 
$U(1) \times SO(3)^2$ broken down to $SO(3)_{\rm V}$
\cite{Volovik}, and in spinor Bose-Einstein condensate 
in cold atom.
}. 
Non-Abelian strings are the most elementary 
objects at the chiral phase transition 
since the $\eta'$ string is a composite state 
made of $N$ of them.  
Although the authors in \cite{Balachandran:2002je} 
constructed solutions with domain walls 
in the presence of the anomaly, 
they did not construct pure string solutions 
in the absence of the anomaly.

In this Letter 
we construct non-Abelian string solutions 
with the axial symmetry 
in the $U(N)_{\rm L} \times U(N)_{\rm R}$ linear sigma model 
in the absence of the anomaly. 
Fine structure of these profile functions will be important
to discuss the detailed dynamics of 
strings at the chiral phase transition.
We find that Nambu-Goldstone bosons 
of ${\bf C}P^{N-1}$ appear 
due to the breaking of $U(N)_{\rm V}$ 
around the fundamental non-Abelian strings,  
as in the case of local strings \cite{Auzzi:2003fs}.
The interaction of  these strings with general
orientations ${\bf C}P^{N-1}$
will be reported in \cite{NNM}.


\section{The Linear Sigma Model}

The Lagrangian of the linear sigma model is given by
\begin{eqnarray}
{\cal L}={\rm tr}\,(\partial_{\mu}\Phi^{\dagger}\partial^{\mu}\Phi
	-m^{2}\Phi^{\dagger}\Phi)  
        -\lambda_{1}[{\rm tr}\,(\Phi^{\dagger}\Phi)]^{2}
	-\lambda_{2}\,{\rm tr}\,[(\Phi^{\dagger}\Phi)^{2}]
	\label{sigma_lag}
\end{eqnarray}
where $\Phi$ is an $N$ by $N$ matrix of complex scalar fields and  
$\lambda_{1}$ and $\lambda_{2}$ 
are the coupling constants \cite{Lenaghan:2000ey}.
The chiral symmetry $U(N)_{\rm L}\times U(N)_{\rm R}$ acts on $\Phi$ as
\begin{eqnarray}
 \Phi \to U_{\rm L} \Phi U_{\rm R}^{\dagger} , \hspace{5mm}
 (U_{\rm L}, U_{\rm R}) \in U(N)_{\rm L}\times U(N)_{\rm R}.
  \label{chiral}
\end{eqnarray}
The static energy density is
${\cal E}={\rm tr}(\partial_{i}\Phi^{\dagger}\partial_{i}\Phi)
	+V $
with the potential 
\begin{eqnarray}
	V(\Phi^{\dagger},\Phi)=m^{2}{\rm tr}(\Phi^{\dagger}\Phi)+\lambda_{1}
	[{\rm tr}(\Phi^{\dagger}\Phi)]^{2}
	+\lambda_{2}{\rm tr}(\Phi^{\dagger}\Phi)^{2} \label{potential}.
\end{eqnarray}
The field equations of $\Phi$ for 
static solutions read
\begin{eqnarray}
 \delta {\cal E}/\delta \Phi^{\dagger}
=[-\nabla^{2}+m^{2}+2\lambda_{1}
	{\rm tr}(\Phi^{\dagger}\Phi)+2\lambda_{2} \Phi \Phi^{\dagger}
  ]\Phi=0 . \label{Phi-eq}
\end{eqnarray}
We consider the parameter region 
\begin{eqnarray}
 m^{2}<0, \hspace{5mm} 
 \lambda_2 > 0, \hspace{5mm}
 N\lambda_{1}+\lambda_{2}>0,   \label{phase-cond}
\end{eqnarray} 
in which the vacuum $\left<\Phi\right>$ is taken to be 
\begin{eqnarray}
	\left<\Phi\right> ={\rm diag}\,(v, \cdots, v), 
\hspace{5mm}
 v\equiv \sqrt{-m^{2}/2(N\lambda_{1}+\lambda_{2})}  
  \label{VEV}
\end{eqnarray}
by using the symmetry (\ref{chiral}). 
The chiral symmetry (\ref{chiral})
is spontaneously broken down to 
a diagonal $U(N)_{\rm V}$; 
\begin{eqnarray}
 \left<\Phi\right> \to U \left<\Phi\right> U^{\dagger} , \hspace{5mm}
 U \in U(N)_{\rm V} .
  \label{U(N)V}
\end{eqnarray}
This breaking 
results in the appearance of 
$N^{2}$ Nambu-Goldstone bosons 
associated with the coset space 
$[U(N)_{\rm L}\times U(N)_{\rm R}] 
/ U(N)_{\rm V} \simeq U(N)_{\rm A}$, 
which are interpreted 
as pions and $\eta'$ meson.
(The latter is massive 
in the presence of the anomaly of $U(1)_{\rm A}$ 
\cite{'t Hooft:1976up}.)


\section{String Solutions}

Since the fundamental homotopy group 
of the chiral symmetry breaking is nontrivial, 
$\pi_1 (G/H) = \pi_1 (U(N)_{\rm A}) \simeq {\bf Z}$, 
there exist vortex-string solutions. 
Let us consider an axially symmetric ansatz 
with multiple winding numbers $n_i \in {\bf Z}$ 
($i=1,\cdots,N$)
\begin{eqnarray}
	\Phi = {\rm diag}\, (f_{1}(r)e^{in_{1}\theta},f_{2}(r)e^{in_{2}\theta}, 
	\cdots , f_{N}(r)e^{in_{N}\theta})\, \label{}
\end{eqnarray}
where $r$ and $\theta$ are radial and angular coordinates of 
two codimensions of vortices. 
We call the solution of this form 
an $(n_{1},n_{2},\cdots ,n_{N})$-vortex.  
The total winding number of $\pi_1$ is $\sum_i n_i$.
With this ansatz, the tension, or 
the static energy per unit length, of strings 
is written as  
\begin{eqnarray}
	E=2\pi \sum_{i=1}^{N} \int \bigg[f_{i}'^{2}
	+\frac{1}{r^{2}}(n_{i}^{2}f_{i}^{2})
	+m^{2}f_{i}^{2} 
       + \lambda_{1}(\sum_{j=1}^{N}f_{j}^{2})
	f_{i}^{2}+\lambda_{2}f_{i}^{4}\bigg]r\,dr \,. \label{}
\end{eqnarray}
The field equation (\ref{Phi-eq}) for $f_{i}$ reduces to
\begin{eqnarray}
  0 =
-f_{i}''-\frac{1}{r}f_{i}'+\frac{n_{i}^{2}}{r^{2}}f_{i}
  +m^{2} f_{i}+2\lambda_{1}(\sum_{j\neq i}^{N}f_{j}^{2})f_{i}
+2(\lambda_{1}+\lambda_{2})f_{i}^{3}\,. \label{gen_field_eq}
\end{eqnarray}
One can see from (\ref{gen_field_eq}) that the asymptotic forms are
\begin{eqnarray}
  f_{i}=a_{n_{i}}r^{n_{i}}+O(r^{n_{i}+2}), 
 \hspace{5mm}
  f_{i}=v+\frac{\alpha_{i}}{r^{2}}+O(r^{-4}) \label{expansions}
\end{eqnarray}
for small and large $r$, respectively, 
where $a_{n_{i}}$ is a shooting parameter determined by the 
boundary condition at infinity, and 
\begin{eqnarray}
&&	\alpha_{i}=-\frac{n_{i}^{2}v}{2pv^{2}+m^{2}}
	-\frac{4\lambda_{1}(N-1)v^{2}}{2pv^{2}+m^{2}}\sum_{j\neq i}^{N}
	\alpha_{j} \,, 
\end{eqnarray}
with $p\equiv (N+2)\lambda_{1}+3\lambda_{2}$.
Using these asymptotic forms, one can compute the leading order of 
the static energy:
\begin{eqnarray}
 E={\rm const.}+2\pi \sum_{i=1}^{N}n_{i}^{2}\,v\,{\rm log}\Lambda\,, 
  \label{energy}
\end{eqnarray}
where $\Lambda$ is an infrared cutoff parameter. 
It is logarithmically divergent as generally expected in the 
theory of global vortices.
This formula implies that 
there exists repulsive force between 
vortices in the same component 
which is well-known in the case of Abelian strings \cite{VS},
while no force is exerted between vortices 
in different components, which is a new feature 
of non-Abelian strings.

Abelian solutions are obtained by the $(1,1,\cdots,1)$ ansatz 
with the tension
\begin{eqnarray}
	E={\rm const.}+2\pi N v{\rm log}\Lambda\,. \label{Abelian-tension}
\end{eqnarray}
The profile is the same with that 
of the Abelian $\eta'$ string \cite{Balachandran:2001qn}. 
The pions do not contribute to this string.

Genuine non-Abelian strings as the minimal solution 
are obtained by the $(1,0,\cdots,0)$ ansatz:  
\begin{eqnarray}
	\Phi_0 (r,\theta)
 ={\rm diag}\,(f(r)e^{i\theta},g(r),\cdots,g(r)) 
	\,. \label{NA-string-ansatz}
\end{eqnarray}
The field equation in (\ref{gen_field_eq}) becomes 
\begin{eqnarray}
	&& \hspace{-3mm} 
         0= -f''-\frac{1}{r}f'+\frac{1}{r^{2}}f 
	+m^{2}f+2\lambda_{1}(N-1)g^{2}f 
        +2(\lambda_{1}+\lambda_{2})f^{3} \label{eq1} , \\
       && \hspace{-3mm} 
        0= -g''-\frac{1}{r}g'+m^{2}g+2\lambda_{1}f^{2}g 
+2[(N-1)\lambda_{1}+\lambda_{2}]g^{3} \,. 
 \nonumber \;\;\;
\end{eqnarray}
The tension  
\begin{eqnarray}
	E= {\rm const.} +2\pi v\,{\rm log}\Lambda 
	\label{NA-tension}
\end{eqnarray}
is $1/N$ of (\ref{Abelian-tension}) for
the Abelian $\eta'$ string, so non-Abelian strings are 
sometimes called $1/N$ (fractional) strings.
It implies that an $\eta'$ string 
can be marginally decomposed into $N$ non-Abelian strings 
without the cost of energy.
The fractional property of $1/N$ strings 
comes from the fact that 
the string (\ref{NA-string-ansatz}) 
winds around a circle in $U(N)_{\rm A}$ generated by
 a linear combination of 
$U(1)_{\rm A}$ and 
$T \sim {\rm diag}(1-N,1,\cdots,1)$ of $SU(N)_{\rm A}$ 
\cite{Balachandran:2002je}.
Therefore $1/N$ strings are composed of both $\eta'$ and pions.

One can see from Eqs.~(\ref{eq1}) that 
when $\lambda_{1}=0$ the profile $g$ is flat, $g =v$,  
and the profile $f$  
is identical to that of the Abelian string.
In this case, a non-Abelian solution 
is simply obtained by 
embedding an Abelian string
to the corner of the $N$ by $N$ matrix 
$\Phi$ in (\ref{VEV}).

Let us discuss zero modes in the presence of 
the string (\ref{NA-string-ansatz}). 
In addition to two translational zero modes,
some internal zero modes 
appear because the solution (\ref{NA-string-ansatz}) 
breaks vacuum symmetry $U(N)_{\rm V}$ given in (\ref{U(N)V})
down to its subgroup $U(N-1) \times U(1)$. 
The Nambu-Goldstone modes for this breaking 
parametrize internal space of the coset space
\begin{eqnarray}
 {\bf C}P^{N-1} = {U(N)_{\rm V} \over U(N-1)_{\rm V} \times U(1)_{\rm V}},
\end{eqnarray}
which is known as the complex projective space.
This was pointed out \cite{Auzzi:2003fs} 
in the case of local non-Abelian strings. 
More explicitly, the most general 
minimal solution is given by $U \Phi_0 (r,\theta) U^{\dagger}$ 
with $\Phi_0$ in Eq. (\ref{NA-string-ansatz}) and $U$ the elements 
of ${\bf C}P^{N-1}$. 
For instance  
$\Phi_0 = {\rm diag}(f e^{i \theta},g)$ for $N=2$ 
is transformed by 
$U = 
\left(\begin{array}{cc}
   \alpha  &  \beta \\
 - \beta^* & \alpha^* 
\end{array}
\right)$ (with $|\alpha|^2 + |\beta|^2 = 1$)
to 
$
\left(\begin{array}{cc}
 |\alpha|^2 f e^{i \theta} + |\beta|^2 g
 &  \alpha \beta   (- f e^{i \theta} + g) \\
  \alpha^* \beta^* (- f e^{i \theta} + g) 
 & |\beta|^2 f e^{i \theta} + |\alpha|^2 g
\end{array}
\right)$. 
This does not depend on the overall phase of 
$\alpha,\beta$, so the space is 
${\bf C}P^{\bf 1} \simeq S^2$.
These modes are non-normalizable 
(in an infinite region) whereas 
the corresponding modes are normalizable in 
the local case \cite{Auzzi:2003fs}.
However in a finite volume 
they are normalizable and are relevant to dynamics.


\section{Numerical computation}

In this section we show 
the numerical results for the fundamental string $(1,0,0)$ and 
a composite string $(1,1,0)$ for $N=3$. 
For that purpose, let us introduce dimensionless quantities,
\begin{eqnarray}
	{\hat x}={\hat m}x \;,\;\;\; {\hat \Phi}=\frac{\sqrt{\lambda_{2}}}{{\hat m}}
	\Phi \,; \hspace{5mm} {\hat m} \equiv \sqrt{-m^{2}}. \label{rescale}
\end{eqnarray}
The Lagrangian~(\ref{sigma_lag}) is rescaled as 
\begin{eqnarray}
	{\hat {\cal L}}={\rm tr}\,(\partial_{\mu}{\hat \Phi}^{\dagger}
	\partial^{\mu}{\hat \Phi}+{\hat \Phi}^{\dagger}{\hat \Phi})
	-\kappa \, [{\rm tr}\,({\hat \Phi}^{\dagger}{\hat \Phi})]^{2}
-{\rm tr}\,({\hat \Phi}^{\dagger}
	{\hat \Phi})^{2} \;\;
\end{eqnarray}
where from Eq. (\ref{phase-cond}) we have 
$\kappa \equiv {\lambda_{1} / \lambda_{2}} 
 > - 1/N$.

\begin{figure}
\includegraphics[height=6.0cm, width=8.5cm]{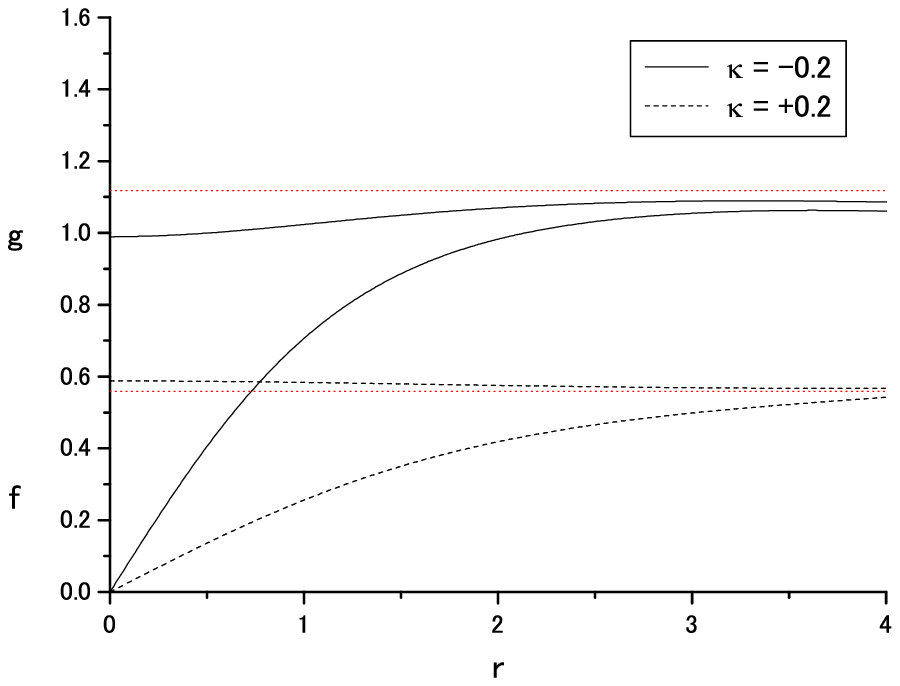}
\includegraphics[height=6.0cm, width=8.5cm]{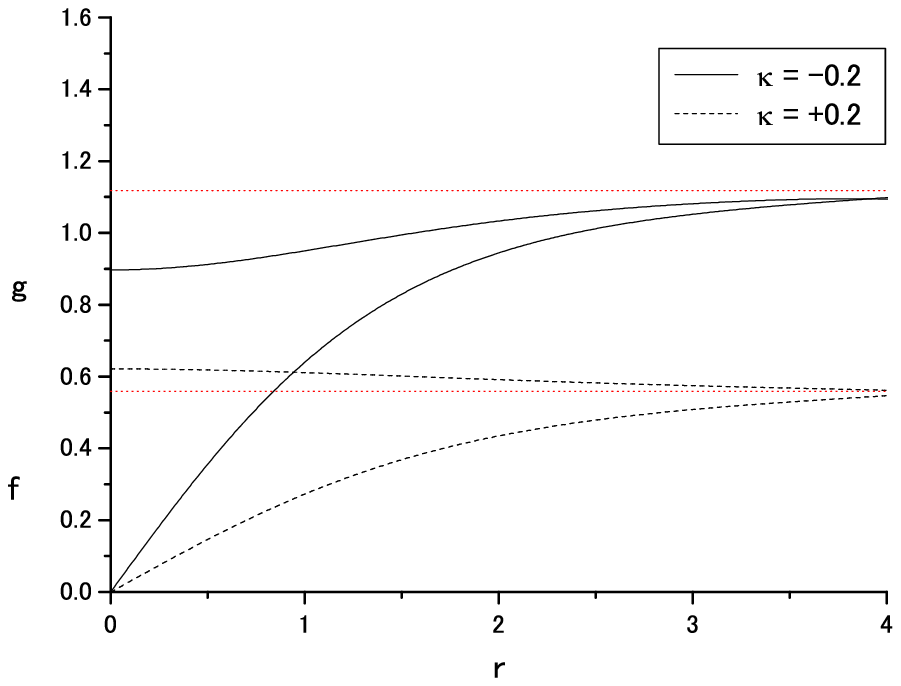}

\hspace{4cm}
{\small (a)}  \hspace{8cm}
{\small (b)}  

\caption{\label{fig} 
The profile functions $f$ and 
$g$ as functions of $r$ 
(with all hats omitted) in the cases of  
(a) $\kappa=-0.2,0.2$ for 
the $(1,0,0)$-vortex, 
and 
(b) $\kappa=-0.2,0.2$
for the $(1,1,0)$-vortex.
The horizontal broken lines denote $v$ to which 
$f$ and $g$ converge.
}
\end{figure}

A. The $(1,0,0)$-vortex.
The ansatz is given by 
\begin{eqnarray}
	{\hat \Phi} (r,\theta)
         =({\hat f}(\hat r) e^{i\theta}, {\hat g}(\hat r), 
	{\hat g}(\hat r))\,,
\end{eqnarray} 
with $\hat r = \hat m r$ and 
$\hat f$ and $\hat g$ rescaled as $\hat \Phi$.
The field equations (\ref{eq1}) 
become
\begin{eqnarray}
	&&0= -{\hat f}''-\frac{1}{{\hat r}}{\hat f}'
	+\frac{1}{{\hat r}^{2}}{\hat f}
	-{\hat f} +4\kappa{\hat g}^{2}{\hat f}
	+2\left(1+\kappa\right){\hat f}^{3} ,
	 \nonumber \\ 
	&&0= -{\hat g}''-\frac{1}{{\hat r}}{\hat g}'-{\hat g}
	+2\kappa {\hat f}^{2}{\hat g}
	+2(1+2\kappa){\hat g}^{3} \label{n31}
\end{eqnarray}
with  prime denoting differentiation with respect to $\hat r$.
The asymptotic forms for small ${\hat r}$ are derived as 
\begin{eqnarray}
	{\hat f}=a_{1}{\hat r}+a_{3}{\hat r}^{3} +O({\hat r}^{4}), \;\;
	{\hat g}=b_{0}+b_{2}{\hat r}^{2}+O({\hat r}^{3}) 
 \label{asympt-numerical}
\end{eqnarray}
with $a_{1}$ and $b_{0}$ are shooting parameters and 
\begin{eqnarray}
	a_{3}=\frac{1}{8}(4\kappa b_{0}^{2}-1)a_{1} ,\;\;
	b_{2}=\frac{1}{4}\left[2(1+2\kappa)b_{0}^{2}
         -1\right]b_{0}\,. \label{}
\end{eqnarray}
Fig. \ref{fig}-(a) shows the profile functions 
for the $(1,0,0)$-vortex. 
The $g$ changes its behavior drastically depending on the 
sign of $\kappa$: 
it is concave, constant ($g=v)$ or convex,
for $\kappa<0$,$\kappa =0$ or $\kappa >0$, respectively.  
For the realistic case of the chiral symmetry breaking 
in the absence of anomaly 
$\kappa$ is negative \cite{Lenaghan:2000ey}.


B. The $(1,1,0)$-vortex. 
The ansatz is given by
\begin{eqnarray}
	{\hat \Phi} (\hat r,\theta)=({\hat f}(\hat r) e^{i\theta}, 
        {\hat f}(\hat r) e^{i\theta}, 
        {\hat g}(\hat r))\, 
\end{eqnarray} 
The corresponding field equations are 
\begin{eqnarray}
	&& 0=-{\hat f}''-\frac{1}{{\hat r}}{\hat f}'
	+\frac{1}{{\hat r}^{2}}{\hat f}
	-{\hat f} +2\kappa{\hat g}^{2}{\hat f}
	+2(1+2\kappa)
	{\hat f}^{3} ,\nonumber \\ 
	&& 0=-{\hat g}''-\frac{1}{{\hat r}}{\hat g}'-{\hat g}
	+4\kappa {\hat f}^{2}{\hat g}
	+2(1+\kappa){\hat g}^{3} . \label{n32}
\end{eqnarray}
The asymptotic forms for small ${\hat r}$ 
are given by (\ref{asympt-numerical}) with 
\begin{eqnarray}
	a_{3}=\frac{1}{8}(2\kappa b_{0}^{2}-1)a_{1},\;\;
	b_{2}=\frac{1}{4}\left[2(1+\kappa)b_{0}^{2}
              -1\right]b_{0} \,.\;\;\;
\end{eqnarray}
The profiles behave similarly to the $(1,0,0)$-vortex profiles 
as shown in 
Fig.~\ref{fig}-(b).

The both figures show that the transverse size of the strings
is of order $\hat x \sim 1$ ($x \sim \hat m^{-1}$) 
as expected. 
From the profile $\hat g$, we confirm that unlike the Abelian case 
the chiral symmetry is recovered only partially in the core of strings 
as speculated in \cite{Balachandran:2002je}; 
$U(1)_{\rm A}$ for the $(1,0,0)$ vortex 
and $U(2)_{\rm A}$ for the $(1,1,0)$ vortex.


\section{Conclusion and Discussion} 
Taking examples of 
$(1,0,0)$ and $(1,1,0)$ vortices in the $N=3$ model,
we have constructed axially symmetric 
non-Abelian string solutions in the linear sigma model. 
The numerical result shows that the profile function $g$
without a winding is concave, flat or convex 
with respect to the radius, depending on 
the sign of the coupling constant $\kappa$.
The profile function $f$ with non-zero winding number  
vanishes at the center of the string as expected.
We have also shown that additional Nambu-Goldstone 
bosons of ${\bf C}P^{N-1}$ appear 
around the fundamental non-Abelian strings, 
as observed in the local strings \cite{Auzzi:2003fs}. 

While Abelian global strings emit the $U(1)$ 
Nambu-Goldstone bosons \cite{VS},  
non-Abelian strings will emit both the $\eta'$ meson and the pions 
when they oscillate. 
We expect that this radiation gives a signal 
which can be detected by a heavy-ion collider 
or in the early Universe.
Interactions between two strings will be described by 
a direct calculation \cite{NNM} or 
by exchange of the pions. 
Interaction between a string and baryons as Skyrmions 
is also an interesting problem.

The obvious extension 
in the context of the chiral symmetry breaking is 
the inclusion of the anomaly term 
\cite{Balachandran:2001qn,Balachandran:2002je} 
and the pion masses. 
In application to cosmology, 
a calculation of the reconnection (intercommutation) rate 
will be important. 
These strings may play some roles in the early Universe 
such as structure formation and the primordial magnetic field 
(see \cite{Brandenberger:1998ew} for the Abelian case). 
The ring of vortex can be a candidate of the dark matter 
(see \cite{Carter:2002te} for the pion string).
The thermal effect and gravity coupling also 
remain as future problems.


\end{document}